\documentstyle[multicol,aps,psfig]{revtex}
\renewcommand{\narrowtext}{\begin{multicols}{2}
\global\columnwidth20.5pc\noindent}
\renewcommand{\widetext}{\end{multicols}
\global\columnwidth42.5pc}
\begin{document}
\draft
\preprint{12 August 2004}
\title{Nuclear Magnetic Relaxation in the Ferrimagnetic Chain Compound
       NiCu(C$_7$H$_6$N$_2$O$_6$)(H$_2$O)$_3$$\cdot$2H$_2$O:
       Three-Magnon Scattering?}
\author{Hiromitsu Hori and Shoji Yamamoto}
\address{Division of Physics, Hokkaido University,
         Sapporo 060-0810, Japan}
\date{Received 12 August 2004}
\maketitle
\begin{abstract}
Recent proton spin-lattice relaxation-time ($T_1$) measurements on
the ferrimagnetic chain compound
NiCu(C$_7$H$_6$N$_2$O$_6$)(H$_2$O)$_3$$\cdot$2H$_2$O
are explained by an elaborately modified spin-wave theory.
We give a strong evidence of {\it the major contribution to $1/T_1$ being
made by the three-magnon scattering rather than the Raman one}.
\end{abstract}
\pacs{PACS numbers: 75.10.Jm, 75.50.Gg, 76.50.$+$g}
\narrowtext

   Low-frequency spin dynamics in magnetic systems is a long-standing
problem and has recently attracted renewed interest due to the significant
progress in designing low-dimensional materials such as chains and
ladders.
Nuclear magnetic resonance (NMR) is a powerful probe to their dynamic
properties.
The nuclear spin-lattice relaxation time $T_1$ is in particular eloquent
of the collective motions of electronic spins and therefore we take a
great interest in microscopically interpreting it.
The spin-wave formalism has played a crucial role in this context.
Van Kranendonk, Bloom \cite{K545}, and Moriya \cite{M23} made their
pioneering attempts to describe $1/T_1$ in terms of spin waves.
Oguchi and Keffer \cite{O405} further developed the spin-wave analysis
considering the three-magnon nuclear relaxation mechanism as well as the
Raman one, whereas Pincus and Beeman \cite{P398} claimed that the
three-magnon process was considerably underestimated in their argument,
revealing further relaxation mechanism.

   The spin-wave excitation energy is usually much larger than the nuclear
resonance frequency and thus the single-magnon relaxation process is
rarely of significance.
The Raman process consequently plays a leading role in the nuclear
spin-lattice relaxation.
Because of the $(4S)^{-1}$-damping factor to the Holstein-Primakoff magnon
series expansion, the multi-magnon scattering is much less contributive
within the first-order mechanism, where a nuclear spin directly interacts
with spin waves through the hyperfine coupling.
However, the second-order mechanism, where a nuclear spin flip induces
virtual spin waves which are then scattered thermally via the four-magnon
exchange interaction, may generally enhance the relaxation rate.
The Pincus-Beeman spin-wave nuclear relaxation theory is thus fascinating
but works only far below the three-dimensional transition temperature.
The conventional spin-wave theory applied to low-dimensional magnets ends
in failure with diverging magnetizations.
In such circumstances, Takahashi \cite{T168} gave a fine description of
the low-dimensional ferromagnetic thermodynamics at low temperatures in
terms of modified spin waves.
His idea of introducing a constraint on the magnetization was developed
for antiferromagnets \cite{T2494,H4769} and ferrimagnets
\cite{Y14008,O8067}.
Random-bond ferromagnets \cite{W014429} and frustrated antiferromagnets
\cite{H2887,C7832,D13821} were also discussed within this renewed
spin-wave scheme.

   The ferrimagnetic modified spin-wave theory is particularly powerful to
investigate both static \cite{Y1024,N214418} and dynamic
\cite{Y157603,H054409} properties.
One-dimensional ferrimagnets have lately attracted much attention
especially in the context of designing molecule-based ferromagnets
\cite{K95}.
A series of bimetallic chain compounds \cite{K782} are typical examples
and some of them were indeed assembled into a ferromagnetic lattice.
Another approach \cite{C1976} consists of bringing into interaction metal
ions and stable organic radicals.
Homometallic systems can exhibit distinct ferrimagnetism of topological
origin \cite{E4466,D83}.
Such synthetic endeavors have stimulated several experimentalists
\cite{F1073,F433} to measure $T_1$ on ferrimagnetic chain compounds.
Thus motivated, here we make a systematic spin-wave analysis on the
nuclear spin dynamics in one-dimensional Heisenberg ferrimagnets,
which has been pending for the past decades without any suitable spin-wave
scheme.
Our goal is to show a strong evidence of {\it the proton spin relaxation
in the title compound being mediated by the three-magnon scattering rather
than the Raman one}.

   First of all our scheme \cite{Y14008} of modifying the spin-wave theory
is distinct from the original idea proposed by Takahashi \cite{T2494}
and Hirsch {\it et al.} \cite{H4769}.
Their way of suppressing the divergent sublattice magnetizations consists
of diagonalizing an effective Hamiltonian with a Lagrange multiplier
included subject to zero staggered magnetization.
The thus-obtained energy spectrum necessarily depends on temperature and
fails to reproduce the Schottky peak of the specific heat \cite{Y064426}.
In order to obtain better thermodynamics, we first diagonalize the
Hamiltonian keeping the dispersion relations free from temperature and
then introduce a Lagrange multiplier in order to minimize the free energy
subject to zero staggered magnetization.
This scheme is highly successful in describing the magnetic susceptibility
as well as the specific heat \cite{Y064426} and therefore guarantees our
exploration of the one-dimensional ferrimagnetic dynamics over a wide
temperature range.

   We consider ferrimagnetic Heisenberg chains of alternating spins $S$
and $s$, as described by the Hamiltonian
\begin{equation}
   {\cal H}
      =\sum_{n=1}^N
       \big[
        J\mbox{\boldmath$S$}_{n}\cdot
         (\mbox{\boldmath$s$}_{n-1}+\mbox{\boldmath$s$}_{n})
       -(g_SS_n^z+g_ss_n^z)\mu_{\rm B}H
       \big].
   \label{E:H}
\end{equation}
Introducing bosonic operators for the spin deviation in each sublattice
via
$S_i^+=(2S-a_{-:i}^\dagger a_{-:i})^{1/2}a_{-:i}$,
$S_i^z=S-a_{-:i}^\dagger a_{-:i}$,
$s_i^+=a_{+:i}^\dagger(2s-a_{+:i}^\dagger a_{+:i})^{1/2}$,
$s_i^z=-s+a_{+:i}^\dagger a_{+:i}$,
and assuming that $O(S)=O(s)$, we expand the Hamiltonian with respect to
$1/S$ as
${\cal H}={\cal H}_2+{\cal H}_1+{\cal H}_0+O(S^{-1})$, where ${\cal H}_i$
contains the $O(S^i)$ terms.
${\cal H}_2\equiv -2SsJN$ is the classical ground-state energy, while
${\cal H}_1$ describes linear spin-wave excitations and is diagonalized
in the momentum space as
\begin{eqnarray}
   &&
   {\cal H}_1
  =-(S+s)JN
   -\big[
     g_S(S+{\textstyle\frac{1}{2}})
    -g_s(s+{\textstyle\frac{1}{2}})
    \big]\mu_{\rm B}HN
   \nonumber\\
   &&\qquad
   +(g_S-g_s)\mu_{\rm B}H
    {\textstyle\sum_k}
    \frac{S+s+(g_S-g_s)\mu_{\rm B}H/2J}{2\omega_k}
   \nonumber\\
   &&\qquad
   +J{\textstyle\sum_k}\omega_k
   +J{\textstyle\sum_k\sum_{\sigma=\pm}}
     \omega_k^\sigma\alpha_{\sigma:k}^\dagger\alpha_{\sigma:k},
\end{eqnarray}
where
$\alpha_{\sigma:k}^\dagger\equiv
 a_{ \sigma:k}^\dagger{\rm cosh}\theta_k
+a_{-\sigma:k}        {\rm sinh}\theta_k$,
provided
${\rm tanh}2\theta_k\,
 =\,2\sqrt{Ss}\cos(k/2)/[S+s+(g_S-g_s)\mu_{\rm B}H/2J]$,
creates a spin wave of ferromagnetic ($\sigma=-$) or antiferromagnetic
($\sigma=+$) aspect \cite{Y13610}, whose energy is given by
$\omega_k^\sigma=\omega_k+\sigma[S-s-(g_S+g_s)\mu_{\rm B}H/2J]$ with
$\omega_k\equiv\{[S+s+(g_S-g_s)\mu_{\rm B}H/2J]^2-4Ss\cos^2(k/2)\}^{1/2}$.
Minimizing the free energy under the condition of zero staggered
magnetization \cite{Y157603}, we obtain the optimum distribution functions
$\bar{n}_k^\sigma\equiv
 \langle\alpha_{\sigma:k}^\dagger\alpha_{\sigma:k}\rangle$.

   The hyperfine interaction is generally expressed as
\begin{eqnarray}
   &&
   {\cal H}_{\rm hf}
   =g_S\mu_{\rm B}\hbar\gamma_{\rm N}I^+
    {\textstyle \sum_n}
     \big(
      {\textstyle\frac{1}{2}}A_n^-S_n^-+A_n^zS_n^z
     \big)
   \nonumber\\
   &&\qquad\ 
   +g_s\mu_{\rm B}\hbar\gamma_{\rm N}I^+
    {\textstyle \sum_n}
     \big(
      {\textstyle\frac{1}{2}}B_n^-s_n^-+B_n^zs_n^z
     \big),
\end{eqnarray}
where $A_n^\sigma$ ($B_n^\sigma$) is the dipolar coupling tensor
between the nuclear and $n$th larger (smaller) electronic spins.
Since ${\cal H}_0$ and ${\cal H}_{\rm hf}$ are both much smaller than
${\cal H}_1$, they act as perturbative interactions to the linear
spin-wave system.
If we consider up to the second-order perturbation with respect to
${\cal V}\equiv{\cal H}_0+{\cal H}_{\rm hf}$, the probability of a nuclear
spin being scattered from the state of $I^z=m$ to that of $I^z=m+1$ is
given by
\begin{equation}
   W=\frac{2\pi}{\hbar}\sum_f
     \Biggl|\Bigl\langle f\Bigl|
      {\cal V}+\sum_{m(\neq i)}
      \frac{{\cal V}|m\rangle\langle m|{\cal V}}{E_i-E_m}
     \Bigr|i\Bigr\rangle\Biggr|^2
     \delta(E_i-E_f),
   \label{E:W}
\end{equation}
\widetext
\vspace*{-4mm}
\begin{figure}
\centerline
{\mbox{\psfig{figure=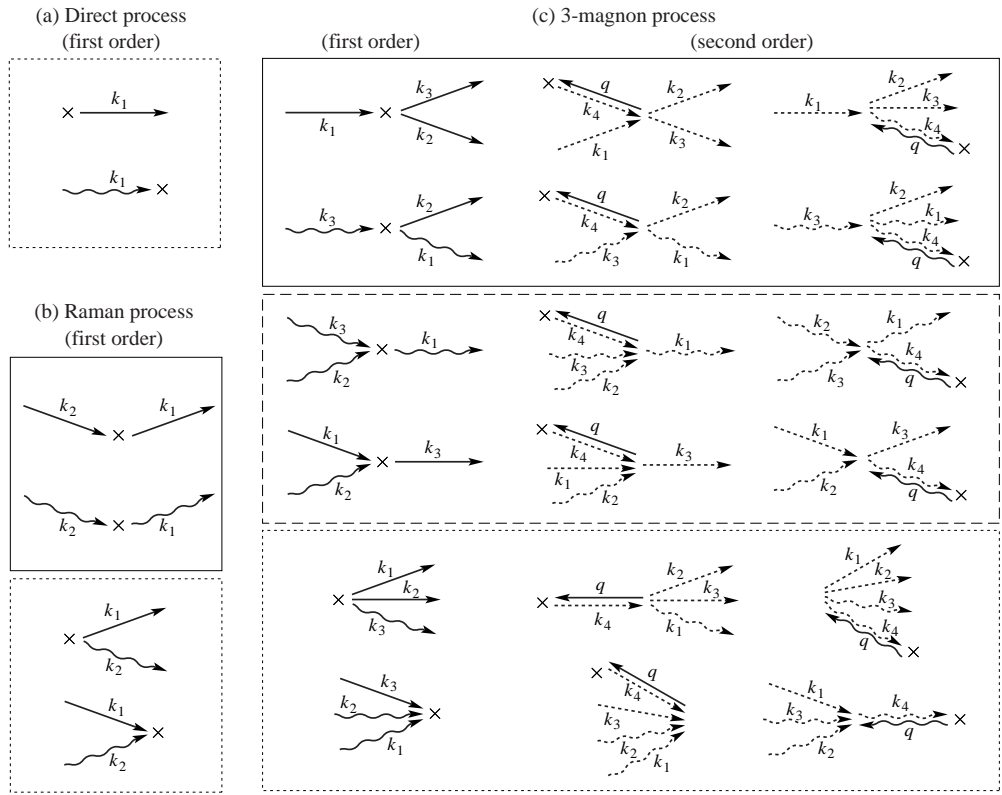,width=132mm,angle=0}}}
\vspace*{2mm}
\caption{Diagrammatic representation of various nuclear spin-lattice
         relaxation processes.
         Solid arrows, designating spin waves which are emitted in the
         first-order processes, induce a nuclear spin flip ($\times$) via
         the hyperfine interaction, while broken arrows, depicting
         four-magnon exchange correlations, thermally scatter the
         first-order spin waves as {\it virtual excitations}, where spin
         waves of ferromagnetic and antiferromagnetic aspect are
         distinguishably drawn by straight and wavy arrows, respectively.
         (a) The first-order direct (single-magnon) relaxation processes;
         (b) The first-order Raman (two-magnon) relaxation processes;
         (c) The first-order and second-order three-magnon relaxation
             processes, where $q=-k_4\equiv k_1-k_2-k_3$, are related to
             each other through nonlinear equations and are therefore
             inseparable.
         Considering the nuclear-electronic energy conservation, processes
         in solid and dotted frames are of great and little significance,
         respectively, whereas those in broken frames are relevant
         according to the constituent spins $S$ and $s$.}
\label{F:diagram}
\end{figure}
\narrowtext
\noindent
where $i$ and $f$ designate the initial and final states of the
unperturbed electronic-nuclear spin system.
Then we find that $T_1=(I-m)(I+m+1)/2W$.
Equation (\ref{E:W}) contains various relaxation processes but their
explicit formulae will be presented elsewhere.
We instead diagrammatically show them in Fig. \ref{F:diagram}.
Due to the considerable difference between the nuclear and electronic
energy scales, $\hbar\omega_{\rm N}\ll J$, the direct process, involving a
single spin wave, is rarely of significance.
Considering further that the antiferromagnetic spin waves are higher in
energy than the ferromagnetic ones, $\omega_k^-<\omega_k^+$, at moderate
fields, the intraband spin-wave scattering dominates the Raman relaxation
rate $1/T_1^{(2)}$, whereas both the intraband and interband spin-wave
scatterings contribute to the three-magnon relaxation rate $1/T_1^{(3)}$.
Within the first-order mechanism, $1/T_1^{(3)}$ is much smaller than
$1/T_1^{(2)}$ \cite{O405}.
However, the first-order relaxation rate is generally enhanced through
the second-order mechanism.
We consider the leading second-order process, that is, the
exchange-scattering-induced three-magnon relaxation, as well as the
first-order process.
The second-order single-magnon and Raman processes, containing three and
two virtual magnons, respectively, are much more accidental
\vspace*{-4mm}
\begin{figure}
\centerline
{\mbox{\psfig{figure=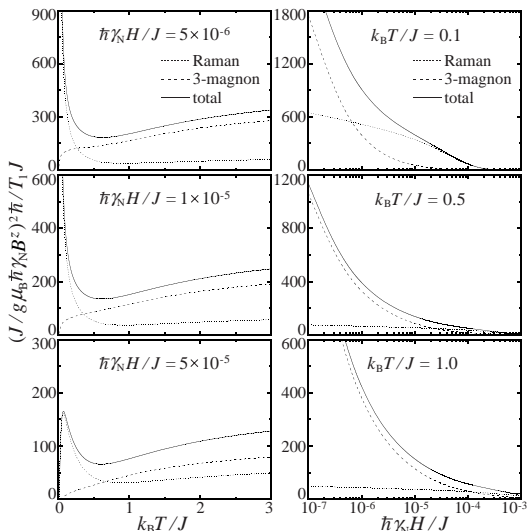,width=70mm,angle=0}}}
\caption{Modified spin-wave calculations of typical temperature (the left
         three) and field (the right three) dependences of the nuclear
         spin-lattice relaxation rate, where $g_S=g_s\equiv g$,
         $A^\tau/B^\tau=1$, and $(B^-/B^z)^2=4$.
         $1/T_1^{(2)}$ and $1/T_1^{(3)}$ are plotted by dotted and broken
         lines, respectively, while $1/T_1^{(2)}+1/T_1^{(3)}\equiv 1/T_1$,
         which is observable, by solid lines.}
\label{F:T1}
\end{figure}
\vspace*{-4mm}
\begin{figure}
\centerline
{\mbox{\psfig{figure=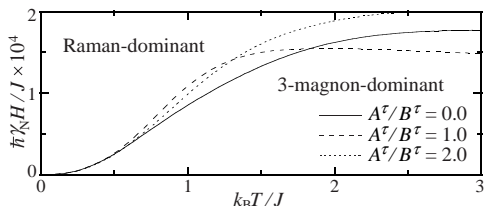,width=64mm,angle=0}}}
\caption{The crossover point as a function of temperature and an applied
         field, where $g_S=g_s$ and $(B^-/B^z)^2=4$.}
\label{F:PhD}
\end{figure}
\noindent
due to the
momentum conservation and much less contributive due to the
$(4S)^{-1}$-damping factor in the Holstein-Primakoff magnon series
expansion.

\vspace*{-2mm}
   We calculate the case of $(S,s)=(1,\frac{1}{2})$, which is relevant to
several major materials \cite{C1976,D83,P138}, assuming that the Fourier
components of the coupling constants have little momentum dependence
\cite{F433} as
$\sum_n e^{{\rm i}kn}A_n^\tau\equiv A_k^\tau\simeq A^\tau$ and
$\sum_n e^{{\rm i}kn}B_n^\tau\equiv B_k^\tau\simeq B^\tau$
($\tau=-,z$).
Figure \ref{F:T1} shows $1/T_1$ as a function of temperature and an
applied field.
The exchange-scattering-enhanced three-magnon relaxation rate generally
grows into a major contribution to $1/T_1$ with increasing temperature and
decreasing field.
As temperature increases, $\bar{n}_k^-$ decreases at $k\simeq 0$ but
otherwise increases \cite{Y2324}.
In one dimension, excitations at $k\simeq 0$ predominate in the Raman
process, while all the excitations are effective in the three-magnon
process.
$1/T_1^{(2)}$ and $1/T_1^{(3)}$ are hence decreasing and increasing
functions of temperature, respectively, unless temperature is so high as
to activate the antiferromagnetic spin waves.
The field dependences of $1/T_1^{(2)}$ and $1/T_1^{(3)}$ are also in
striking contrast.
At moderately low temperatures and weak fields,
$\hbar\omega_{\rm N}\ll k_{\rm B}T\ll J$, we find that
$1/T_1^{(2)}\propto
 e^{-(g_S+g_s)\mu_{\rm B}H/2k_{\rm B}T}
 K_0(\hbar\omega_{\rm N}/2k_{\rm B}T)$,
where $K_0$ is the modified Bessel function of the second kind and
behaves as
$K_0(\hbar\omega_{\rm N}/2k_{\rm B}T)
 \simeq 0.80908-{\rm ln}(\hbar\omega_{\rm N}/k_{\rm B}T)$.
Thus the field dependence of $1/T_1^{(2)}$ is initially logarithmic
and then turns exponential with increasing field.
Equation (\ref{E:W}) claims that $1/T_1^{(3)}$ is much less analyzable but
should exhibit much stronger power-law diverging behavior with decreasing
field.
Therefore, the three-magnon relaxation process predominates over the Raman
one at weak fields.

\vspace*{-2mm}
   In Fig. \ref{F:PhD} we plot {\it the crossover points} on which
$1/T_1^{(2)}=1/T_1^{(3)}$.
A Raman-to-three-magnon crossover may generally be detected with
increasing temperature and decreasing field.
The ferrimagnetic nuclear spin-latticerelaxation is sensitive to another
adjustable parameter $A^\tau/B^\tau$, that is, the location of the probe
nuclei.
At the special location of
$A^\tau/B^\tau\sim(d_s/d_S)^3\simeq(S/s)^\sigma$, where $d_S$ ($d_s$) is
the distance between the nuclear and larger (smaller) electronic spins,
the $\sigma$ excitation mode hardly mediates the nuclear spin relaxation
\cite{Y2324}.
For $(S,s)=(1,\frac{1}{2})$, the lower-lying ferromagnetic spin waves
are almost invisible to the nuclear spin located as $A^\tau/B^\tau$
\vspace*{-5mm}
\begin{figure}
\centerline
{\mbox{\psfig{figure=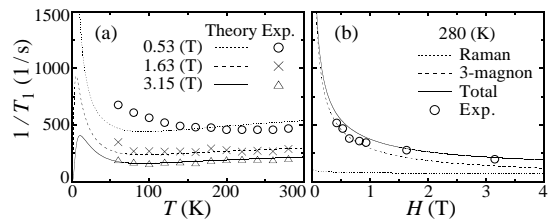,width=72mm,angle=0}}}
\caption{Proton spin relaxation-time measurements on
         NiCu(C$_7$H$_6$N$_2$O$_6$)(H$_2$O)$_3$$\cdot$2H$_2$O [24]
         compared with our theory.
         (a) $1/T_1$ as a function of temperature at various values of
             an applied field;
         (b) $1/T_1$ as a function of an applied field at $280\,\mbox{K}$,
             where $1/T_1^{(2)}$ and $1/T_1^{(3)}$ are also plotted by
             dotted and broken lines, respectively.}
\label{F:experiment}
\end{figure}
\noindent
$\simeq 1/2$
and therefore its relaxation rate stays extremely small.
Any $T_1$ measurements should be performed away from such magic points.

   We are further excited to compare our theory with recent experimental
findings.
Fujiwara and Hagiwara \cite{F433} measured $T_1$ for proton nuclei in the
bimetallic chain compound
NiCu(C$_7$H$_6$N$_2$O$_6$)(H$_2$O)$_3$$\cdot$$2$H$_2$O \cite{P138}
comprising ferrimagnetic chains with alternating octahedral Ni$^{2+}$ and
square-pyramidal Cu$^{2+}$ ions.
The measured susceptibility \cite{H2209} suggests that
$J/k_{\rm B}\simeq 121\,\mbox{K}$, $g_S=2.22$, and $g_s=2.09$.
Comparative measurements on the D$_2$O-substituted samples \cite{F433}
indicate that protons relevant to the $T_1$ findings are located in close
vicinity to Cu spins.
Then, considering that the dipolar coupling strength is in proportion to
the inverse cubic distance, we may set the coupling constants for
$A^\tau/B^\tau=0$.
Conditioning further that $B^z=1.37\times 10^{20}\,\mbox{T}^2/\mbox{J}$
and $(B^-/B^z)^2=5$, which can be consistent with the crystalline structure
\cite{P138}, we plot calculations together with the observations in
Fig. \ref{F:experiment}.
The ferromagnetic and antiferromagnetic spin waves contribute different
temperature dependences to $1/T_1$ and give the decreasing and then
increasing behavior in Fig. \ref{F:experiment}(a).
Considering that there may be larger uncertainty in the experimental
findings for $1/T_1$ at lower temperatures and weaker fields \cite{F433},
the theoretical and experimental findings are in good agreement and the
slight discrepancy between them may be attributable, for instance, to weak
momentum dependence of $B_k^\tau$ and the protons of wide distribution.
Figure \ref{F:experiment}(b) more impressively demonstrates the
relevance of the three-magnon scattering to the proton spin relaxation.
{\it The strong field dependence can never be explained by the Raman
process}.
Since $1/T_1^{(3)}$ within the first-order mechanism stays much smaller
than the observations, {\it the exchange-scattering-induced three-magnon
process is essential in interpreting such accelerated relaxation}.
We are eager to have reliable observations at lower temperatures and
weaker fields.
We call for more extensive NMR measurements using as probes
$^1$H, $^{63}$Cu, and $^{55}$Mn nuclei on the family material
MnCu(C$_7$H$_6$N$_2$O$_6$)(H$_2$O)$_3$$\cdot$$2$H$_2$O \cite{P138}
as well as that of present interest.

   There exist pioneering $T_1$ measurements on the layered ferromagnet
CrCl$_3$ \cite{N354} and the coupled-chain antiferromagnet
CsMnCl$_3\cdot$2H$_2$O \cite{N5325}, which give evidence of the relevant
three-magnon scattering.
However, they are both, in some sense, {\it classical} findings under the
existing three-dimensional long-range order.
Without any reliable spin-wave formulation in one dimension, no author
has explored {\it quantum} ferrimagnetic dynamics with particular
interest in multi-magnon scattering beyond the Raman mechanism.
The present calculation is {\it the first evidence of the three-magnon
scattering dominating the one-dimensional nuclear spin relaxation} and
motivate extensive $T_1$ measurements on various one-dimensional quantum
ferrimagnets.


\widetext
\end{document}